=============================================================================             
  PAPER FOLLOWS
  --------------------------------------------------------------------------
  \documentstyle[aps,twocolumn,prl,epsf]{revtex}
  
\begin{document}
  
  \newcommand{\be}{\begin{equation}}
  \newcommand{\ee}{\end{equation}}
  \newcommand{\beq}{\begin{eqnarray}}
  \newcommand{\eeq}{\end{eqnarray}}
  \newcommand{\lsim}{\:\raisebox{-0.5ex}{$\stackrel{\textstyle<}{\sim}$}\:}
  \newcommand{\gsim}{\:\raisebox{-0.5ex}{$\stackrel{\textstyle>}{\sim}$}\:}
  
  \draft
  
  \twocolumn[\hsize\textwidth\columnwidth\hsize\csname @twocolumnfalse\endcsname
  
  \title{On the origin of the Non-Fermi Liquid Behavior of $SrRuO_{3}$.}
  
  \author{M. S. Laad and E. M\"uller-Hartmann}
  
  \address{Institut f\"ur Theoretische Physik, Universit\"at zu K\"oln, 77 
  Z\"ulpicher Strasse, 50937 K\"oln, Germany \\
  }
  
  \date{\today}
  
  \maketitle
  
  \widetext

  \begin{abstract}
    Motivated by the unusual features observed in the transport properties of the
  ferromagnetic "bad metal" $SrRuO_{3}$, we construct a model incorporating 
  essential features of the 
  realistic structure of this nearly cubic material.  In particular, we show 
  how the $t_{2g}$ orbital {\it orientation} in the perfectly cubic structure
   determines the peculiar structure of the 
  hybridization matrix, and demonstrate how the local 
  non-Fermi liquid features arise 
  when interactions are switched on.  We discuss the effects of the slight 
  deviation from the cubic structure qualitatively. 
   The model provides a consistent explanation of the features observed recently in the optical response of $SrRuO_{3}$. 
  \end{abstract}

  ]
  
  \narrowtext
  
     Strongly correlated electronic systems continue to present new surprises 
  in
  condensed matter physics.  In particular, $d$-electron systems exhibit
  fascinating phenomena, from unconventional high-$T_{c}$ superconductivity 
  to
   colossal magnetoresistance [1].  These dramatic behaviors are believed to 
  be
  associated with the fact that the important electronic states are often
  intermediate between the itinerant and localized limits, necessitating the
  development of new ideas [2].
  
     Ruthenates constitute a class of $d$-band oxides occuring in both 
  layered
  [3,4] as well as nearly cubic forms.  The electronic states responsible for 
  both
  conduction and magnetism (or superconductivity) are associated with bands
  which involve $Ru$-4d orbitals hybridized with the $O$-2p states.  The 
  cubic
  material $SrRuO_{3}$ exhibits a transition from a high-$T$ paramagnetic
  metallic to a ferromagnetic metallic state below $T_{c}^{FM}=150 K$ [5].
  State-of-the-art bandstructure calculations do produce this basic feature 
  [6].
  However, photoemission [7] measurements reveal a much reduced spectral 
  density
  at the Fermi surface, compared to the results from a bandstructure 
  calculation.  Moreover, the authors of [7] do not observe a clear
  separation between the "coherent" and incoherent parts of the spectral 
  function, suggesting that the low-energy spectral weight may have a 
  significant incoherent contribution.
  The mass enhancement factor deduced from this is  
  in excellent agreement 
  with
  that extracted from the low-$T$ specific heat, showing the importance of
  correlation effects.  More unusual features are observed in transport 
   studies.
  At low $T$, a resistivity minimum is observed even as the resistivity itself 
  is
  very low [8,9,10].  At higher $T>T_{c}^{FM}$, $\rho_{dc}(T)$ increases linearly with 
  $T$,
  passing through the Ioffe-Regel limit [8] without saturation.  It exhibits a 
  kink at $T_{c}^{FM}$ and continues to increase up to $1000$K.  In other 
  words, one is
  dealing with a situation where $k_{F}l \simeq 1$ (already for $T \simeq 500$K)
  , suggesting the 
  inapplicability
    of Boltzmann equation approaches for transport.  In fact, this similarity 
  with other "bad metals", as well as with dilute ferromagnetic alloys [8], suggests that a local picture (as opposed to a band description) for transport is
  required for $SrRuO_{3}$.  Near $T_{c}^{FM}$, 
   anomalous
  critical behavior in $\rho_{dc}(T)$ has been found by Klein {\it et al.} 
  [10].
  That this is not related to magnetic critical behavior is shown by the fact
  that the magnetization data are well fit with conventional universal  
  critical
  exponents.  This line of thinking is further reinforced by a recent study [10] 
  probing the deviations from Matthiessen's rule (DMR) in $SrRuO_{3}$ and $CaRuO_{3}$.
  The latter is not a ferromagnet down to the lowest temperatures.  Interestingly, however, the similarity in the radiation-induced change in the resistivity for {\it both} samples indicates that the observed DMR is not caused by magnetic 
  ordering.   More recently, Kostic {\it et al.} [11] have studied the 
  infrared
  conductivity of $SrRuO_{3}$ from $50 cm^{-1}$ to $40000 cm^{-1}$ at 
  temperatures
    ranging from $40$K to $300$K.  Quite astonishingly, $\sigma(\omega) 
  \simeq
  \omega^{-\alpha}$ with $\alpha=1/2$ deep into the ferromagnetic phase 
  ($T=40$K)
  instead of the $\omega^{-2}$ form characteristic of Fermi liquids.  This 
  form
  is close to the phenomenological fit used for doped cuprates [12].
  Motivated by the strong correlation features observed experimentally, 
  Ahn {\it et al.} [13] attempted to model the optical spectrum in terms of the
  response of a one-band Hubbard model in $d=\infty$.  However, the low-$T$ 
  response in the $d=\infty$ Hubbard model is that of a correlated Fermi liquid,
  and Ahn {\it et al.} themselves point out the importance of extra scattering 
  mechanisms to understand the non-FL fall-off in $\sigma(\omega)$.  In 
  particular, they speculate that factors like the $p-d$ hybridization and its 
  interplay with Mott-Hubbard physics might provide some insight into the 
  physics of ruthenates. 
     Given the similarity in their optical response, one might be tempted to look 
  for a possible explanation in terms of theories proposed to understand the 
  non-FL metallic phase of cuprates.  In sharp contrast to the cuprates, however, $SrRuO_{3}$ is not close to any
  quantum critical point.  There are no antiferromagnetic spin fluctuations,
  neither does it undergo a Mott transition.  Actually, it is not clear whether
  $SrRuO_{3}$ is close to a Mott transition.  The FM spin fluctuations cannot
  be the cause of the non-FL features, since these features
   are observed well below
  $T_{c}^{FM}$.
  The scattering
  rate extracted from an extended "Drude" fit increases {\it linearly} with
  frequency, and remains high above $T_{c}^{FM}$.  At high $T>T_{c}^{FM}$, $\sigma(\omega)$ 
  even appears to increase with frequency in the far-infrared, reminiscent of
  strongly disordered systems.  However, the nature of such strong scattering
  and, in particular, its microscopic origin, is completely unclear, given 
  that
  this system has negligible extrinsic disorder [8].
    One is led, therefore, to search
  for the origin of the (intrinsic) strong scattering in the peculiarities of
  the material itself.
  
    A proper understanding of the anomalies should, in the final analysis, be
  linked to the peculiarities in the local quantum chemistry (or the basic 
  electronic structure) of this material.  In this connection, it is interesting
  to notice that Cox {\it et al.} [14] have reported {\it three} flat bands 
  crossing the Fermi surface: one $e_{g}$ band between $\Gamma$ and $X$, and
  two $t_{2g}$ bands between the $X$ and $M$ points in the Brillouin zone. 
  The microscopic origin of these features are, however, completely obscured in
  the numerics.
  
    The appearance of a ferromagnetic metallic state at low $T$ in $SrRuO_{3}$ is
  understood as follows.
    In $SrRuO_{3}$, the crystal field breaks the first Hund's rule and the four $Ru$ d electrons then partially fill the
  $t_{2g}$ states, giving rise to a local spin $S=1$ at each site, consistent with
   the measured magnetization ($M \simeq 1 \mu_{B}$) at low $T$.  The $t_{2g}^{4}$ configuration implies a three-fold orbital degeneracy since the fourth electron can go into any of the three $t_{2g}$ orbitals.  If this was the  
 whole story (i.e, if we ignored the $O-2p$ bands), one would expect $SrRuO_{3}$ to be a $S=1$, antiferromagnetic
  Mott insulator (probably with some kind of orbital ordering).  We claim
   that this orbital degeneracy in the perfectly cubic system is the origin of 
  the strong intrinsic scattering observed experimentally (see below).
    Bandstructure calculations [14] as well as photoemission measurements, 
   however, point to the important role of self-doping in this
  material, introducing a small concentration of holes into the $O-2p$ band via 
  charge transfer processes, which put a small concentration of additional 
  electrons into the $t_{2g}$ orbitals.  The
  Ru $4d_{xy, yz, zx}$-O $2p$ hybridization then 
   drives the resulting mixed-valent
  system metallic and ferromagnetic via the double-exchange mechanism.
  
     In this letter, we propose a microscopic model for $SrRuO_{3}$ which 
  ties
  together (i) the local quantum chemistry, (ii) the relevant interactions, 
  and
  (iii) the observed non-Fermi liquid anomalies within one approach.
  In particular, we show how the $t_{2g}$ 
  orbital orientation determines the peculiar
  structure of the hybridization matrix (see below), and how the NFL behavior
  arises (within a single-impurity picture) 
   when local coulomb interactions are switched on.
  
     We start by deriving the hybridization (Ru $4d_{xy, yz, zx}$-O $2p$) 
  matrix
  for the nearly cubic structure of $SrRuO_{3}$.  To begin with, we shall 
  assume
  a perfect cubic structure, and consider the effects of the slight 
  distortion later in the paper.
  
     Considering the local orbital structure of the $Ru-O$ octahedron, the 
  relevant electronically active states involve the hybridization of the triply
  degenerate $Ru-4d_{xy,yz,zx}$
  and $O-2p_{x,y,z}$ states.  In this situation, the hopping part is
  written as
  
  \be
  \nonumber
 H_{hop}=-t\sum_{{\bf l}\sigma}[d_{xy{\bf l}\sigma}^{\dag}P_{{\bf l}\sigma} + c.t. +h.c.]
 \ee
 where $P_{{\bf l}\sigma}=(p_{y,{\bf l
 }+{\bf x/2},\sigma}-p_{y,{\bf l}-{\bf x/2},\sigma}+p_{x,{\bf l}+{\bf y/2},\sigma
 }-p_{x,{\bf l}-{\bf y/2},\sigma})$ and 
  $p_{\alpha,{\bf l}\pm\beta/2}$ means the $p_{\alpha}$ orbital placed at 
 the center of the bond(s) leaving site $i$ in the $\pm\beta$ direction. 
   In eqn.(1), $c.t.$ (cyclic terms) means terms which arise by a 
 cyclic permutation of $x,y,z$.  Notice that the $d_{xy}$ orbital {\it does not} 
 hybridize with
  $p_{z,l\pm\alpha/2}$ and $p_{\alpha,l\pm z/2}$
  ($\alpha=x,y$) because of symmetry reasons.  It is 
 precisely
   this peculiarity resulting from the local quantum chemistry of 
 $SrRuO_{3}$
 which, together with the orbital degeneracy mentioned above, is responsible for its anomalous behavior, as we show below.
 
   The hopping part of the hamiltonian eqn.(1) describes a planar hybridization of each of the $t_{2g}$ 4d-orbital of $Ru$ with a hybridizing $O$-2p (Wannier)
  orbital
 in close formal analogy to the $3d_{x^{2}-y^{2}}-2p_{\sigma}$ hybridization in
 the $CuO_{2}$ planes in cuprates.
 
   In order to make the peculiarity mentioned above more transparent, we Fourier transform the hopping part (1), giving,
 
 \beq
 \nonumber
 H_{hop}(t_{2g}) &=& -2it\sum_{{\bf k}\sigma}[d_{xy{\bf 
 k}\sigma}^{\dag}(s_{x}p_{yx{\bf k}\sigma}+s_{y}p_{xy{\bf 
 k}\sigma}) \\ & &   
  + c.t. + h.c.] \;. 
 \eeq
 using the short notations, 
 $s_{\alpha}=\sin(k_{\alpha}/2)$ ($\alpha=x,y,z$), and defining the 
 hybridizing (Wannier) orbitals, $w_{yz{\bf 
 k}}=-i(s_{y}p_{zy{\bf k}}+s_{z}p_{yz{\bf k}})/f_{yz}$, $w_{xz{\bf 
 k}}=-i(s_{z}p_{xz{\bf k}}+s_{x}p_{zx{\bf k}})/f_{zx}$, and $w_{xy{\bf 
 k}}=-i(s_{x}p_{yx{\bf k}}+s_{y}p_{xy{\bf k}})/f_{xy}$.  Writing 
 $f_{\alpha\beta}=(s_{\alpha}^{2}+s_{\beta}^{2})^{1/2}$  
 and going back to the Wannier representation we get, finally,
 
 \beq
 \nonumber
 H_{hop}(t_{2g}) &=& -t\sum_{<lm>\sigma}[T_{yz}(m)d_{yzl\sigma}^{\dag}w_{yz,l+m\sigma} \\ & & + c.t. + h.c.] \;.
 \eeq
 with $T_{\alpha\beta}(m)=\langle 2(s_{\alpha}^{2}+s_{\beta}^{2})^{1/2}e^{i{\bf k}.{\bf 
 m}}\rangle_{\bf k}$.
   Thus, in the perfect cubic
   symmetry, the
 bandstructure is purely two-dimensional in each direction solely as a consequence of the
 orbital {\it orientation}.
 
    We now consider the interaction part of the Hamiltonian.
 The local, intra-orbital Hubbard interaction is large, as can be seen from 
 the
 satellite feature in photoemission being situated $\simeq 5$eV below the 
 Fermi
 energy.  We can thus roughly estimate $U_{dd} \simeq 5$eV (the results 
 obtained below are insensitive to its actual value).  The Hund's rule
 coupling is of the order of magnitude of the exchange splitting ($\simeq 
 0.5$eV).  The nearest neighbor $Ru-O$ interaction, $U_{pd}$, can be estimated
 to be $U_{pd} \simeq (e^{2}/a\epsilon) \simeq 1.5-2$eV.  Setting $U_{dd}\rightarrow\infty$, we write the interaction term as [15],
 \be
 H_{int}=U_{pd}\sum_{<lm>\sigma}n_{ld\sigma}(n_{l+m,w,\sigma'}^{x}+n_{l+m,w,\sigma'}^{y}+n_{l+m,w,\sigma'}^{z}).            
 \ee
 
 Thus, the total Hamiltonian reads 
 \be
 H=H_{hop}(t_{2g})+H_{int}.
 \ee
 
   The peculiar feature coming from the local quantum chemistry of $SrRuO_{3}$ 
 is now clear.  The $w_{z\alpha}$ do not hybridize with the $d_{xy}$ orbitals,
 but interact with a strength $U_{pd}$, $w_{x\alpha}$ do not hybridize with the $d_{yz}$, and the $w_{y\alpha}$ do not hybridize with the $d_{zx}$.  Hence, for 
 example,
 the term $U_{pd}\sum_{<lm>\sigma}n_{ldxy\sigma}n_{l+m,w,\sigma'}^{z}$ provides a 
 nonhybridizing (screening) channel that strongly scatters the $w_{z}$-band carriers.  Within the impurity approximation, 
  local non-Fermi liquid behavior [16] now arises from the competition of
 this term, which drives the system to a non-FL state characterized by X-ray edge
  singularities, and the hybridization, which would drive the system to a FL
 fixed point if the nonhybridizing channel were absent.  The situation is somewhat reminiscent of the picture proposed in the context of
 marginal Fermi liquids in cuprates [16].  The origin of the local non-FL behavior is
 quite different, however, in our case.  In what follows, we show
  that the true low-energy response is indeed power-law in nature, as seems to be indicated by optical
 measurements.
 
   Let us first consider the simplified situation when the screening channel 
 (the term $U_{pd}\sum_{<lm>\sigma}n_{ldxy\sigma}n_{l+m,w,\sigma'}^{z}$ )
 is inoperative.  In this case, the hybridization term drives the system to a 
 local Fermi liquid fixed point, as is understood from the observation that 
 $t_{pd}$ is a relevant variable [16].  A detailed solution of this simplified
 model in $d=\infty$ would lead to the same conclusion.  This problem can then
 be mapped onto the problem of quasiparticles hopping in an effective medium
 determined by a complex self-energy, $\Sigma(\omega)$, having the correct Fermi 
 liquid form at low energy, 
 
 \be
 \Sigma_{w\alpha\beta}(\omega)=-(U_{pd}\rho_{0})^{2}[\frac{4\omega}{\pi}+i\frac{\pi\rho_{0}}{2}(\omega^{2}+\pi^{2}T^{2})]
 \ee
 so that the effective model {\it without} the
 screening channel can be written as,
 \be
 H_{eff}^{0}=\sum_{k\sigma\alpha\beta}[\epsilon_{\alpha\beta}({\bf k})+\Sigma_{\alpha\beta}(\omega)]
 w_{\alpha\beta{\bf k}\sigma}^{\dag}w_{\alpha\beta{\bf k}\sigma}
 \ee
 
   The presence of the screening channel modifies the effective hamiltonian, 
 which now reads,
 
 \beq
 \nonumber 
 H_{eff} & = & H_{eff}^{0} + 
 U_{pd}\sum_{l{\bf k}{\bf k'},\alpha\beta\gamma\sigma\sigma'}n_{ld\alpha\beta\sigma}w_{\gamma{\bf k}\sigma'}^{\dag}w_{\gamma{\bf k'}\sigma'}
 \eeq  
 where $\alpha\beta=xy, yz, zx$ in pairs, and 
 $\gamma=z$, when $\alpha\beta=xy$, etc.  This is recognizable as a model where
 the {\it absence} of the $d_{\alpha\beta}-w_{\gamma}$ hybridization results in
 X-ray edge-like singularity right at the Fermi surface, resulting in breakdown of 
 FLT.   We again emphasize that this strong scattering off the "localized"
 $w_{\gamma}$ orbital results from the intrinsic feature of the quantum chemistry of $SrRuO_{3}$, namely, from the orbital degeneracy in the $t_{2g}$ sector and the
  peculiar structure of the hybridization matrix derived above,
  and is not related to extrinsic disorder-induced localization.
   
   To obtain an analytic insight into the origin of the anomalous non-FL 
 features in the optical response, we make some simplifications which will 
 not qualitatively modify the conclusions.  The dispersive band just below the 
 Fermi level is approximated by $\epsilon_{\alpha\beta}(k)=A-Bk^{2}$ for small
 $k$.  Furthermore, we assume that the lattice form factors are assumed 
 constant, since they are expected to be smooth functions of $k$.  The above
 approximations remove the dependence of two-particle quantities on $k$ as well
 as on the orbital indices, and the problem is thereby transformed into that
 of a band of $w$ electrons scattered locally by a "localized" $d$ hole 
 potential right at the Fermi surface.  As is known [17], this problem is 
 exactly soluble in the impurity limit;  we work only in the metallic phase, and so 
  choose a lorentzian unperturbed DOS.  As long as there is no band-splitting, we expect this simplification to hold qualitatively.
 Within the impurity approximation, the
 local self-energy {\it without} the screening term has the Fermi liquid form ; we ignore it in this analysis, as it will not qualitatively modify the
 following physics.  
   The $w$ electron DOS 
 is now given by [18],
 
 \be
 \rho_{w}(\omega)=\frac{1-n_{d}}{\omega^{2}+D^{2}} + \frac{n_{d}}{(\omega-U_{pd})^{2}+D^{2}}
 \ee
 where $D$ is the $w$ bandwidth.  The "localized" $d$-hole spectral function is
 [18]
 
 \be
 \rho_{d}(\omega) \simeq (1-n_{w})\frac{\theta(\omega-\mu)}{|\omega-\mu|^{1-\eta}} + n_{w}\frac{\theta(\omega-U_{pd}-\mu)}{|\omega-U_{pd}-\mu|^{1-\eta}}
 \ee
 where $\pi\eta=tan^{-1}(\pi\rho_{w}(0) U_{pd})$ is the $s$-wave scattering phase shift of 
 the $w$ electrons.  Here, $\rho_{w}(0)$ is the density of states of the $w$ 
 fermions at the Fermi surface, and equals $1/D\pi$ for the lorentzian used 
 here.  The optical response has contributions from both of the 
 above.  The most singular intraband contribution to $\sigma(\omega)$ results
 from $\rho_{d}(\omega)$ above.  Indeed, by a simple scaling argument [19],
 
 \be
 \sigma_{intra}(\omega)=\frac{const}{(i\omega)^{1-2\eta}}
 \ee 
   With a plausible ratio $U_{pd}/D \simeq 1$, we obtain $\eta=1/4$ and the 
 optical conductivity at small $\omega$, $\sigma_{intra}(\omega)=C\omega^{-1/2}$
 as indeed observed experimentally. 
 
   The above relation holds up to an upper energy cut-off of order of $D$.
 There will also be a non-singular contribution from the $w$-DOS at higher 
 energies.
   It is obvious that this is very different from a FL-like description, where
 $\sigma(\omega) \simeq \omega^{-2}$.     
 
   The above holds for a perfectly cubic lattice.  In practice, in $SrRuO_{3}$,
 the lattice is slightly distorted from the perfect cubic structure.  Given 
 this, one expects that the three-fold orbital degeneracy in the $t_{2g}$ sector
 will be slightly lifted, and that the hybridization matrix will contain a small term
 corresponding to $d_{xy}-w_{z}$ hybridization.  Indeed, we expect that the
 strict orthogonality of the $p_{x,y,z}$ orbitals in the perfect cubic structure 
  is violated in the presence of a slight lattice distortion, resulting in a 
 term $t_{pd}'\sum_{il\sigma}(d_{ixy\sigma}^{\dag}p_{i+l/2,z\sigma}+h.c.)$ in 
 eqn.(1).
 We expect the non-FL behavior (the infrared singularity of the X-ray edge type)
 to be cut off by a low-energy quasicoherent scale related to $t_{pd}'$.  At 
 sufficiently low-$T$, (within the local approach) $t_{pd}'$ will become 
 relevant [20], making the system scale to a low-$T$ FL fixed point.  Given
 the small distortion observed in $SrRuO_{3}$, we expect this scale to be quite
 small, consistent with the $T_{0}=40K$ [21] below which a $T^{2}$ component in
 the dc resistivity is indeed observed.  A quantitative estimate of $T_{0}$ 
 is, however, out of the scope of the present work.     
 
   In conclusion, we have developed a model incorporating the basic local 
 quantum chemical aspects of the structure of the nearly cubic ferromagnetic 
 "bad metal" $SrRuO_{3}$.  We have pointed at the three-fold degeneracy of the
 $t_{2g}^{4}$ configuration, as well as at the peculiar features in the hybridization 
 matrix, whose structure is controlled by the orbital {\it orientation} in the
 $t_{2g}$ sector, 
  and is effectively two-dimensional in the perfectly cubic 
 structure.  Finally, we have shown how a non-Fermi liquid metallic state is
 generated within an impurity picture when interactions are switched on, and argued how a small distortion 
 cuts off the low-energy singularity, generating a low energy coherence scale. 
   The model explains the non-FL features observed in optical spectra in a consistent way.

 {\bf ACKNOWLEDGEMENTS}  MSL wishes to thank the Deutsche Forschungsgemeinschaft
 (DFG) for financial support.

 {\bf REFERENCES}
 
 [1] M. Imada, A. Fujimori and Y. Tokura, Rev. Mod. Phys. {\bf 70}, 1039 (1998).
 
 [2] A. Georges, {\it et al.}, Rev. Mod. Phys. {\bf 68}, 13 (1996).

 [3] Y. Maeno, Physica C {\bf 282-287}, 206 (1997).
 
 [4] Tamio Oguchi, Phys. Rev. B {\bf 51}, 1385 (1995).

 [5] C. B. Eom {\it et al.}, Science {\bf 258}, 1766 (1992).
 
 [6] P. B. Allen {\it et al.}, Phys. Rev. B {\bf 53}, 4393 (1996).

 [7] K. Fujioka {\it et al.}, Phys. Rev. B {\bf 56}, 11, 6380 (1997). 
 [8]  T. H. Geballe {\it et al.}, J. Phys. Cond. Matt. {\bf 8}, 10111 (1996).
 
 [9] G. Cao {\it et al.}, Phys. Rev. B {\bf 56}, 321 (1997).
 
 [10] L. Klein {\it et al.}, Phys. Rev. Lett. {\bf 77}, 2774 (1996).
  
 [11] P. Kostic {\it et al.}, Phys. Rev. Lett. {\bf 81}, 2498 (1998).
 
 [12] P. W. Anderson, Phys. Rev. B {\bf 55}, 11785 (1997).
 see also, D. van der Marel, Phys. Rev. B {\bf 60}, R765 (1999).

 [13] J. S. Ahn {\it et al.}, Phys. Rev. Lett {\bf 82}, 5321 (1999).
 
 [14] P. A. Cox {\it et al.}, J. Phys. Cond. Matt. {\bf 16}, 6221 (1983).
 
 [15]  
 Terms like $U_{pd}\sum_{lmm'\sigma\sigma'\alpha\beta\gamma}
 n_{ld\sigma}w_{l+m\sigma'}^{\dag}w_{l+m'\sigma'}$ will be generated in 
 general.  However,
 these have no qualitative effect on the discussion in the text, since they do 
 not modify the main feature coming from the competition of the hybridization and the presence of the screening channel in our model.
 [16] See the Reference [2], and 
 references therein relating to the "solution" of the multiband Hubbard model
 with extra screening channels in $d=\infty$.
 
 [17] G. M. Zhang and L. Yu, Phys. Rev. Lett. {\bf 81}, 4192 (1998).

 [18] K. D. Schotte {\it et al.}, Phys. Rev. {\bf 182}, 479 (1968).
 
 [19] P. W. Anderson, Phys. Rev. B {\bf 55}, 11785 (1997).
 
 [20] The effect of $t_{pd}'$ is to block the singularity caused by strong
 scattering of the "light" $p$ carriers by the "infinitely heavy" $d$-hole
 (caused by $U_{pd}\sum_{<ij>\sigma}n_{id\sigma}^{xy}n_{jp\sigma'}^{z}$); the
 problem then becomes equivalent to allowing the infinitely heavy $d$-hole to 
 acquire a heavy, but finite mass, leading to recoil and destruction of the
 low-energy singular features below a temperature (energy) scale related to
 $t_{pd}'$.  See E. M\"uller-Hartmann {\it et al.}, Phys. Rev. B {\bf 3}, 1102 (1971). 

 [21] A. P. MacKenzie {\it et al.}, Phys. Rev. B {\bf 58}, R13318 (1998). 
 \end{document}